\def \cm{~\rm{cm}}
\def \s{~\rm{s}}
\def \km{~\rm{km}}

\def \g{~\rm{g}}

\def \erg{~\rm{erg}}

\def \yr{~\rm{yr}}
\def \pc{~\rm{pc}}
\def \kpc{~\rm{kpc}}

\documentclass[12pt,preprint]{aastex}
\usepackage{natbib}

\shorttitle{Pairs of bubbles}
\shortauthors{Soker}

\begin{document}

\title{BUBBLES IN PLANETARY NEBULAE AND CLUSTERS OF GALAXIES:
SLOWLY PRECESSING JETS}

\author{Noam Soker\altaffilmark{1}}

\altaffiltext{1}{Department of Physics, Technion, Haifa 32000, Israel;
soker@physics.technion.ac.il.}

\begin{abstract}

I derive the condition for narrow jets with varying axis, e.g., precessing jets,
to inflate more or less spherical (fat) bubbles in planetary nebulae and
clusters of galaxies.
This work follows a previous work dealing with wide jets, i.e., having
a wide opening angle.
The expressions derive here are qualitatively and quantitatively similar
to the conditions for inflating fat bubbles by non-precessing wide jets.
This follows the similar physical cause of inflating fat bubbles, which is that
the jet deposits energy inside the bubble.
Fat bubbles in planetary nebulae (and similar stellar systems) and in clusters
of galaxies, are likely to be formed by wide jets, precessing jets, or other
jets whose axis is not constant relative to the medium they expand into.

\end{abstract}


\section{INTRODUCTION} \label{sec:intro}

{\it Chandra} X-ray observations of clusters of galaxies reveal
the presence of X-ray-deficient bubbles in the inner regions of
many cooling flow clusters of galaxies, groups of galaxies, and
elliptical galaxies, e.g., Hydra A (McNamara et al.\ 2000),
Perseus (Fabian et al.\ 2000, 2003), A 2597 (McNamara et al.\ 2001),
RBS797 (Schindler et al.\ 2001), Abell~4059 (Heinz et al.\ 2002),
MS 0735.6+7421 (McNamara et al. 2005),
Abell~2052, (Blanton et al. 2003), HCG 62 (Vrtilek et al.\ 2002), and
M84 (Finoguenov \& Jones 2001).
Radio emission inside these bubbles indicate that they were inflated
by jets launched by the central active galactic nuclei (AGNs).
The bubbles and jets play a key role in the energy and mass cycle of the
intra-cluster medium (ICM) in these objects
(e.g., Br\"uggen et al.\ 2002; Br\"uggen \& Kaiser 2002;
Heinz \& Churazov 2005; see review by Peterson \& Fabian 2006).

In a previous paper (Soker 2003; see astro-ph version of Soker \& Bisker 2006
for more detail and images) I pointed out an interesting and
not trivial similarity in the morphology and some non-dimensional
quantities between pairs of X-ray-deficient bubbles in clusters of
galaxies and pairs of optical-deficient bubbles in planetary
nebulae (PNs). Examples of PNs with nice pairs of bubbles are the
Owl nebula (NGC 3587; PN G148.4+57.0: Guerrero et al.\ 2003),  Cn
3-1 (VV 171; PN G038.2+12.0; Sahai 2000), and Hu 2-1 (PN G051.4+09.6;
Miranda et al.\ 2001).

{{{  It is imperative to emphasize here that I try to explain the formation
of a low density pair of bubbles with a narrow waist between them.
These bubbles have more or less diffuse radio emission and are
different from radio bubbles with strong radio hot-spots, e.g.,
e.g., 3C285 (Saslaw et al. 1978) and 3C219 (Perley et al. 1980),
that Scheuer (1982) accounted for by precessing jets;
the bubbles of 3C219 don't coincide with X-ray cavities (Brunetti et al. 1999).
The narrow waist between the bubbles I try to explain make them fundamentally
different from cases where there is only one prolate spheroidal cavity,
such as in Cygnus A (Wilson et al. 2006).
For that, the cocoon dynamics studied by Begelman \&  Cioffi (1989) is not
directly applicable to the present study.
}}}

The basic condition for a jet to inflate a fat, more or less spherical,
bubble is that the bubble encloses the jet's head.
Namely, the jet's head will reside inside the bubble during the inflation phase.
In Soker (2004) I derived the condition for a jet having a constant propagation axis.
In that case the condition for the jet's head to be inside the bubble is that the expansion
velocity of the bubble's radius, $v_b$, be larger than the propagation of the jet's head
along its axis $v_b \ga v_h$.
For PNs and other similar stellar systems, the approximate condition on
the half opening angle of a non-precessing jet to inflate a fat bubble is
\begin{equation}
\alpha \gtrsim 40 ^\circ
\left( \frac {v_j}{400 \km \s^{-1}} \right)^{1/10}
\left( \frac {\dot M_j} {0.01 \dot M_s} \right) ^{3/10},
\label{pn1}
\end{equation}
where $v_j$ is the velocity of material inside the jet, $\dot M_j$ is the
mass flow rate inside the jet, and $\dot M_s$ is the mass loss rate into the
spherically symmetric slow, $v_s \simeq 10 \km \s^{-1} \ll v_j$, wind into which the
jet expands (for more detail see section 2 in Soker 2004).

The condition for a jet to inflate a fat bubble in a cluster can be also
written as a condition on the half opening angle $\alpha$ (eq. 14 in Soker 2004).
Narrower jets will have to propagate to larger distances in order to inflate a fat bubble.
It is possible to write the condition on the distance from the center along
the jet's axis where the jet is capable to inflate a bubble
\begin{eqnarray}
z_{\rm bub}({\rm wide}) \ga 10
\left( \frac {\alpha}{65 ^\circ} \right)^{-1}
\left( \frac {\dot E_j} {10^{45} \erg \s^{-1}} \right)^{3/10}
\left( \frac {{v_j}}{10^4 \km \s^{-1}} \right)^{-1/2}
\nonumber \\ \times
\left(\frac {\rho_c}{10^{-25} \g  cm^{-3}} \right)^{-3/10}
\left(\frac {\tau}{10^7 \rm yr} \right)^{2/5} \kpc ,
\label{clust1}
\end{eqnarray}
where $\rho_c$ is the density of the ICM at the location of the bubble,
$\tau$ is the age of the bubble,
and the jet is a non-relativistic jet, with a kinetic power of
$\dot E_j = \dot M_j v_j^2/2$.

The results of Vernaleo \& Reynolds (2006) agree with this estimate.
For a jet power of $\dot E_j=9.8 \times 10^{45} \erg \s^{-1}$, a speed of
$10,500 \km \s^{-1}$, and a half opening angle of $\alpha=15 ^\circ$,
they find the jet to inflate a bubble at a distance of
$z_{\rm bub} \sim 400 \kpc$ and time of $\tau \sim 2 \times 10^8 \yr$.
In many cooling flow clusters the bubbles are much closer to the center,
implying a larger opening angle (Soker 2004), or the destruction of the jet by
a relative motion of the ICM (Loken et al. 1995; Soker \& Bisker 2006), or other
types of changes in the jet's axis, such as precession (Soker 2004), or a random
change (Heinz et al. 2006).

Motivated by the discussions and talks in a recent meeting on cooling
flows in galaxies and clusters of galaxies (Pratt et al. 2007),
I study in the present paper the condition for inflating fat bubbles by very
narrow precessing jets.
Namely, I consider the limit case where the axis returns to a previous direction
over an average time much longer than the inflation time of the bubble.
In Soker (2004) I considered the other extreme, where the jet's
axis does not change and the jet is wide.
The cases in between narrow precessing jets studied here and wide jets
studied in Soker (2004) should be explored by numerical simulations.
{{{  Narrow jets that precess very fast will return
to a previous direction several times during the inflation phase.
They actually have the same physics as wide jets, and will inflate bubbles
according to the conditions in (Soker 2004).
In the present paper I derive the condition on
slowly precessing jets that don't return to a previous directions.
Such jets can leave a signature on the bubble, e.g., one side is
brighter than the other. This could be the case in MS 0735.6+7421
(McNamara et al. 2005).  }}}
In Sec. 2 the general condition is derived.
In Sec. 3 this condition is applied to PNs and similar stellar systems,
and in Sec. 4 to clusters of galaxies.
A short summary is in Sec. 5.
As this paper is a continuation to the study conducted in Soker (2004),
only essential background material is given here. More details and the motivation
for studying inflation of fat bubbles both in PNs and clusters are in earlier
papers (Soker 2003, 2004; Soker \& Bisker 2006).

\section{THE CONDITION FOR INFLATING A FAT BUBBLE BY A SLOWLY PRECESSING JET}
\label{jets}

By precession I refer also to other types of jet's motion, e.g., due to
the relative motion of the ICM and the cD galaxy, or a stochastic variation
of the jet's axis.
I emphasize again that I take the limit case where the jet's axis dose not
return to the same direction.

A narrow jet will expand through the medium (slow AGB wind in PNs and
the ICM in clusters) until the supply of jet's material ends.
Let us consider a conical narrow jet having a half opening angle $\alpha$, and
let the direction of the jet's axis change at a rate $\dot \theta$
(radians per second).
Where ever the jet center passes, material were flowing for
a time of $\alpha/\dot \theta$ before the jet axis was along that direction,
and jet's material will continue to flow along that direction for a time
$\alpha/\dot \theta$ after that.
During that time the jet's head will propagate to a distance
\begin{equation}
z_{\rm bub}({\rm precess})=\int_0^{t_t} v_h(t) dt,
\label{zbub1}
\end{equation}
where $v_h$ is the jet's head propagation speed through the medium and
\begin{equation}
t_t \equiv \frac {2 \alpha}{\dot \theta}.
\label{tp1}
\end{equation}
Therefore, the inflation distance of the bubble is determined by the
opening angle and the precession rate, as well as the jet's speed.

A condition for the inflation of a more or less spherical (fat) bubble is
that the energy be injected inside the already existing bubble; the pressure of the
hot bubble's interior will act then to make the bubble spherical.
If during the bubble age $\tau$ the jet's head moves a transverse distance
$D_b$, the condition reads
\begin{equation}
D_b(\tau) \la R_b (\tau),
\label{cond1}
\end{equation}
{{{  where $R_b(\tau)$ is the bubble's radius at age $\tau$, taken according
to Castor et al. (1975).
Again, I consider slowly precessing jets, and not fast precessing jets, that
have physics similar to wide jets. }}}
Let us consider a precessing jet with the jet's axis angle to the precession
axis being $\beta$, and the precession period $T_p$, implying that
$\dot \theta = 2 \pi \sin \beta/T_p$.
The maximum distance between two points along the jet's head is the diameter of the
precessing jet's head $D_b \simeq 2 z_{\rm bub} \sin \beta$, reached after a time $T_p/2$.
For $\tau= T_p/2$ condition (\ref {cond1}) can be written
as $ z_{\rm bub} \dot \theta T_p/\pi \la R_b(T_p/2)$.
Under the assumptions made here (Soker 2004) the speed of the bubble surface is
$v_b =\dot R_b =(3/5) R_b/\tau$, which allows us to approximate condition (\ref {cond1})
in a general way
\begin{equation}
\dot \theta z_{\rm bub} \eta \la v_b (\tau).
\label{cond2}
\end{equation}
For a precessing jet and for $\tau=T_p/2$ we found above $\eta=6/5 \pi=0.38$.
For randomly moving jet axis (as discussed by Heinz et al. 2006), the
jet's head moves less from its original position, and $\eta$ is smaller.

To inflate a fat bubble by a jet with a moving axis, e.g. precessing jet,
close to the center the rate of change in axis' angle, $\dot \theta$, is constraint
from below by condition (\ref{zbub1}).
For the bubble to be spherical more or less (a `fat' bubble) $\dot \theta$ is
constraint from above by condition (\ref{cond2}), or more precisely
by condition (\ref{cond1}).
In Sec. (3) we derived this condition for PNs and related stellar binary systems,
and in Sec. (4) for cooling flow clusters of galaxies.

\section{JETS IN STELLAR SYSTEMS}
\label{pnjet}
The flow structure is of a jet flowing into the dense circumstellar
material, which is the expanding AGB (or a similar giant star) wind.
For a narrow jet, $\alpha \ll 1$, the distance the jet reaches after a time $t$
is (Soker 2004)
\begin{equation}
z_{\rm bub} =v_h t_t=
\left( \frac{\dot M_j v_j }{\dot M_s v_s} \right)^{1/2} \frac{2}{\alpha}v_s t_t,
\label{zbpn1}
\end{equation}
where $v_s$ is the velocity of the slow wind, assumed to be spherically symmetric,
into which the jet material expands with a speed $v_j$, and $v_h$ is the speed
of the jet's head.
$\dot M_s$ and $\dot M_j$ are the mass loss rates into the slow wind and
one jet, respectively; both are constant, and defined positively.
Substituting typical values for PNs, and $t=t_t=2 \alpha/\dot \theta$ from
equation (\ref{tp1}), yields
\begin{equation}
z_{\rm bub}= 8 
\times 10^{16} \left( \frac {\dot M_j} {0.01 \dot M_s} \right) ^{1/2}
\left( \frac {v_j}{400 \km \s^{-1}} \right)^{1/2}
\left( \frac {v_s} {10 \km \s^{-1}} \right)^{1/2}
\left( \frac {\dot \theta} {{\rm rad}/1000 \yr} \right)^{-1} \cm .
\label{zbpn2}
\end{equation}
This is the inflation region of the bubble, which under the assumptions used here
does not depend on the opening angle of the jet.

If the jet precess on a time scale of  several thousand years the jet
will reach a distance of $\la 0.1 \pc$, and can in principle inflate a bubble
close to the center. However, there are two other conditions
in PNs.
First, the radiative cooling time of the post-shock jet's material
must be long compare to the inflation time. This in principle implies
a jet's speed of $v_j \ga 200 \km \s^{-1}$ (Soker 2004), depending on the
distance from the center where the bubble is formed.

Second,  condition (\ref{cond2}) should be met for a fat bubble to be formed.
The expansion velocity if the jet's head in stellar winds is constant,
and using expression (\ref{zbpn1}) for $z_{\rm bub}$ in condition (\ref{cond2})
with $t=t_t$ gives
\begin{equation}
2 \alpha \eta  v_h \la v_b (\tau).
\label{cond22}
\end{equation}
The constraint on a wide jet expanding along a constant axis to
inflate a fat bubble is (Soker 2004) $v_h \la v_b (\tau)$. It is
clear that the condition for a {{{  slowly}}} precessing jet is
easier to meet since $2 \alpha \eta <1$ for narrow jets. Rapidly
precessing jet's (large $\dot \theta$) are more efficient to
inflate fat bubbles near the center, but they will not reach a
large distance from the center (eq. \ref{zbpn2}). In any case, the
basic physics behind the condition for inflating fat bubbles in
the two extremes (constant jet's axis and slowly precessing very
narrow jet) is similar, leading to similar expression.

Substituting the expression for $v_b$ (eq. 4 of Soker 2004) and $v_h$
as given in equation (\ref{zbpn1}) above in condition (\ref{cond22})
results in a condition for the inflation of a fat bubble
{{{  by a slowly precessing jet}}}
\begin{eqnarray}
{z_{\rm bub}} \ga 1  
\times 10^{16} \left( \frac{\eta}{0.3} \right) ^{5/2}
\left( \frac {v_j}{400 \km \s^{-1}} \right)^{1/4}
\left( \frac {\dot M_j} {0.01 \dot M_s} \right) ^{3/4}
\left( \frac {v_s} {10 \km \s^{-1}} \right)^{3/4}
\left( \frac {\tau} {1000 \yr } \right) \cm.
\label{condf1}
\end{eqnarray}
Under the assumption that the jet's axis never repeats its direction and
the jet is very narrow, this condition does not depend on the opening angle
of the jet. It does depend on $\dot \theta$ via the dependence of the
inflation distance $z_{\rm bub}$ on $\dot \theta$ (eq. \ref{zbpn2}).

The expression for a wide jet having a constant propagation axis
(rearranged version of eq. 6 in Soker 2004) is similar, but a term
$(\alpha/37^\circ)^{-1}$ (written as $(\beta/0.1)^{-1/2}$ in Soker 2004) replaces
$(\eta/0.3)$ here, and the coefficient is $8 \times 10^{16} \cm$
instead of $1 \times 10^{16} \cm$.
Namely, the condition for inflating a fat bubble with a wide jet (or rapidly precessing
narrow jet) is similar to that with a slowly precessing jet,
both qualitatively and quantitatively
(for a wide jet with $\alpha \sim 1 \sim 60^{\circ}$).
For further elaboration on the properties of bubbles in PNs see Soker (2004).

The term $z_{\rm bub}$ can be eliminated from equations (\ref{zbpn2}) and (\ref{condf1}),
yielding the following condition for inflating a fat bubble in stellar winds by
{{{  slowly }}} precessing narrow jets
\begin{eqnarray}
0.14  \la
\left( \frac{\eta}{0.3} \right) ^{-5/2}
\left( \frac {v_j}{40v_s} \right)^{1/4}
\left( \frac {\dot M_j} {0.01 \dot M_s} \right) ^{-1/4}
\left( \frac {\dot \theta} {{\rm rad}/1000 \yr} \right)^{-1}
\left( \frac {\tau} {1000 \yr } \right)^{-1} .
\label{condf2}
\end{eqnarray}

\section{JET PROPAGATION IN CLUSTERS OF GALAXIES}
\label{cljet}

There are several significant differences between bubble inflation
in PNs, or stellar systems in general, and in clusters (Soker 2004).
(1) In clusters the thermal pressure of the ambient
gas is non-negligible. However, following previous papers (Soker 2003, 2004)
I neglect this pressure.
(2) In clusters the ambient medium does not flow outward.
(3) The ICM density profile in the inner regions of clusters, $\rho_c(r)$,
 is much shallower than that of the slow wind from stars.
 I take it as a constant here.
(4) The inflating jets in clusters may be relativistic, and the magnetic
pressure inside the bubble can be large.
(5) In clusters the bubbles can be observed as they form, unlike in PNs,
where they are observed long after the jets have ceased
(old bubbles may be observed in clusters$-$termed ghost-bubbles$-$as in
the Perseus cluster; Fabian et al. 2000).
However, as argued previously (Soker 2003, 2004), these don't prevent a similar
bubble-formation mechanism in PNs and clusters.

By neglecting the magnetic pressure inside the jet and relativistic effects,
hence $\dot E_j = \dot M_j v_j^2/2$, the jet's head speed through the ICM is (e.g., Krause 2003)
\begin{equation}
v_h \simeq \frac{1}{\alpha} \left(\frac {2 \dot E_j}{\pi v_j \rho_c} \right)^{1/2} \frac {1}{z}.
\label{vhcl}
\end{equation}
where, as before, $z$ is the distance from the cluster center along the jet axis.
Substituting equation (\ref{vhcl}) in equation (\ref{zbub1}) and solving gives
\begin{equation}
z=2 v_ht .
\label{vh2}
\end{equation}
Using this relation in condition (\ref{cond2}) with $t=t_t={2 \alpha}/{\dot \theta}$
for $z=z_{\rm bub}$ gives
\begin{equation}
4 \alpha \eta  v_h \la v_b (\tau).
\label{cond23}
\end{equation}
This is similar, up to a factor of 2, to equation (\ref{cond22}).
The condition for a wide jet expanding along a constant axis to inflate
a fat bubble is (Soker 2004) $v_h \la v_b (\tau)$.
As in PNs, it is clear that the condition for a precessing jet is easier to meet
since $4 \alpha \eta <1$ for narrow jets.
Again, the basic physics behind the condition for inflating fat bubbles
in the two extremes, constant axis of a wide jet {{{  (or rapidly precessing jet) }}}
and slowly precessing very narrow jet, is similar, leading to similar expressions.

Substituting the expression for the expansion speed of the bubble front
(eq. 11 from Soker 2004) and equation (\ref{vhcl}) above in equation
(\ref{cond23}) gives the condition for inflating a fat bubble by a {{{  slowly }}}
precessing jet close to the center in clusters of galaxies
\begin{eqnarray}
z_{\rm bub}({\rm wide}) \ga 13.6    
\left( \frac{\eta}{0.3} \right)
\left( \frac {\dot E_j} {10^{45} \erg \s^{-1}} \right)^{3/10}
\left( \frac {{v_j}}{10^4 \km \s^{-1}} \right)^{-1/2}
\nonumber \\ \times
\left(\frac {\rho_c}{10^{-25} \g  cm^{-3}} \right)^{-3/10}
\left(\frac {\tau}{10^7 \rm yr} \right)^{2/5} \kpc ,
\label{clust2}
\end{eqnarray}
Comparing this condition with that for wide jets having a constant axis direction
(Soker 2004; eq. \ref{clust1} here)
we find that the basic physics is the same, qualitatively and quantitatively.
Similar to precessing jets in stellar systems, under the assumption that the
jet's axis never repeats its direction and the jet is very narrow, condition
(\ref{clust2}) does not depend on the opening angle of the jet.
It does depend on $\dot \theta$ via the dependence of the
inflation distance $z_{\rm bub}=2 v_h (z_{\rm bub}) t_t$ on $\dot \theta$.
Explicitly this is
\begin{eqnarray}
z_{\rm bub}= 2
 \left(\frac {2 \dot E_j}{\pi v_j \rho_c} \right)^{1/4}
\dot \theta ^{-1/2} =
5.8  
\left( \frac {\dot E_j} {10^{45} \erg \s^{-1}} \right)^{1/4}
\left( \frac {v_j}{10^4 \km \s^{-1}} \right)^{-1/4}
\nonumber \\ \times
\left(\frac {\rho_c}{10^{-25} \g  cm^{-3}} \right)^{-1/4}
\left( \frac {\dot \theta} {{\rm rad}/10^6 \yr} \right)^{-1/2} .
\label{zbubcl}
\end{eqnarray}

Combining equations (\ref{zbubcl}) and (\ref{clust2}) the condition for inflating a bubble
(at the distance $z_{\rm bub}$) becomes
\begin{eqnarray}
2.3 \la    
\left( \frac{\eta}{0.3} \right)^{-1}
\left( \frac {\dot E_j} {10^{45} \erg \s^{-1}} \right)^{-1/20}
\left( \frac {v_j}{10^4 \km \s^{-1}} \right)^{1/4}
\nonumber \\ \times
\left(\frac {\rho_c}{10^{-25} \g  cm^{-3}} \right)^{1/20}
\left( \frac {\dot \theta} {{\rm rad}/10^6 \yr} \right)^{-1/2}
\left(\frac {\tau}{10^7 \rm yr} \right)^{-2/5} .
\label{conclf}
\end{eqnarray}
The qualitative difference between equation (\ref{conclf}) and
equation (\ref{condf2}) is that in stellar wind the ambient density
falls as $r^{-2}$, which is $z^{-2}$ along the jet axis, while in
clusters of galaxies close to the center the density
falls very slowly.


\section{DISCUSSION AND SUMMARY}
\label{sec:conclusion}

The main goal here was to derive approximate conditions to inflate more or
less spherical (fat) bubbles by {{{   slowly}}} precessing jets, or other jets with
varying direction of propagation, close to their origin.
I examined the condition for stellar winds, e.g., PNs, and for clusters
of galaxies, emphasizing the similar physics between these two classes of objects.
Narrow jets with changing axis' direction reach a maximum distance
given by equation (\ref{zbpn2}) for stellar winds, and equation (\ref{zbubcl})
for clusters.

To inflate a fat bubble the jet's head should inject energy
inside the bubble. The condition is given by equation (\ref{cond1}), or
in approximate form by equation (\ref{cond2}).
The conditions for inflating a fat bubbles in stellar wind is
given by equation (\ref{condf1}), or equation (\ref{condf2}).
The conditions for inflating a fat bubbles in clusters of galaxies
is given by equation (\ref{clust2}), or equation (\ref{conclf}).

{{{  Rapidly precessing narrow jets have the same physics as than of wide opening
angle jets with a constant axis.
Narrow jets with precessing speed in between the slow jets studied here and
rapidly precessing jets will inflate an azimuthally elongated structure,
rather than one fat bubble, and no clear dense waist will exist between
the two sides.
}}}

The expressions in stellar winds and clusters of galaxies have the
same basic physics, with similar dependence on the rate of change
of the jet's axis direction $\dot \theta$, age of bubble $\tau$, and the
jets parameters.
Indeed, as was argued before (Soker 2003, 2004; Soker \& Bisker 2006)
there are similarities in morphology, some non-dimensional quantities,
and formation mechanism, between X-ray-deficient bubbles in clusters
of galaxies and the optical-deficient bubbles in PNs.

The expressions derive here are qualitatively similar to the condition
for inflating fat bubbles by jets with wide opening angle (Soker 2004).
The expression are quantitatively similar when the non-precessing wide jet has
an opening angle of $\alpha \sim 1 \sim 60^\circ$.
The similarity in the physics of inflating bubbles by wide jets (or
rapidly precessing jets), slowly precessing jets,
and motion of the ICM was discussed previously (Soker 2004).

{{{  In some cases where a fat bubble is inflated close to the origin, the jet's axis
changes direction on time scales short relative to the inflation time
(but not very short), and it will be difficult to observe the precession itself
(very fast precession will behave like a wide-opening angle jet).
In cases of very slow precession (low value of $\dot \theta$) the jet will propagate to
a large distance, and will change direction over a long time such that the
precession can be observed; in some of these cases bubbles will be formed, in other not.
An example of a bubble pair at a large distance from the origin where the precession
is observed is the galaxy cluster MS 0735.6+7421 (McNamara et al. 2005), for which
Pizzolato \& Soker (2005) suggested that a binary black hole system, similar to binary
stellar systems in PNs, is responsible for the precession.
There are many PNs that show precession signatures, e.g., IPHAS PN-1 for
a recent one (Mampaso et al. 2006).
Bubbles formed by AGN jets with a long precession period are observed also in
Seyfert galaxies (e.g., Markarian 6; Kharb et la. 2006).
These cases, as well as cases where new jets have a different direction to that
of old jets, e.g., the cluster  RBS797 (Gitti et al. 2006), show that precession
is quite common.
However, in the present paper the main goal was to understand the formation of of a
pair of fat bubbles with a waist between them,
where in most cases the precession is not expected to be observed directly.  }}}

Although the  bulk motion of the ICM relative to the jet, e.g., shear and rotation,
can also lead to the inflation of fat bubbles (Loken et al.\, 1995; Heinz et al. 2006),
this is not a necessary ingredient, as in PNs in general such motions do not exist,
but still fat bubbles are formed. Wide jets and precessing (or randomly moving)
jets (Soker 2004) are more likely to be the cause in most cases.
Bulk motion of the ICM in clusters, or of the wind in stellar system,
will cause a large scale departure from axisymmetry (Soker \& Bisker 2006).

\acknowledgements
This research was supported by the Asher Fund for Space Research at the
Technion.

\end{document}